\documentclass[11pt]{article}
\usepackage[margin=1in]{geometry}
\usepackage{amsmath,amssymb}
\usepackage{graphicx}
\usepackage{booktabs}
\usepackage{hyperref}
\usepackage[section]{placeins}
\usepackage{float}
\usepackage{authblk}
\usepackage[round]{natbib}
\usepackage{xcolor}
\usepackage{microtype}
\usepackage{afterpage}
\usepackage{longtable}
\usepackage{array}
\newcolumntype{L}[1]{>{\raggedright\arraybackslash}p{#1}}

\hypersetup{colorlinks=true,linkcolor=blue,citecolor=blue,urlcolor=blue,
pdftitle={Backspace as a Natural Experiment: An Accelerated Failure Time Model of Selective Post-Error Motor Impairment in Parkinson's Disease},
pdfauthor={Navin Bondade}}

\title{\textbf{Backspace as a Natural Experiment: An Accelerated Failure Time Model of Selective Post-Error Motor Impairment in Parkinson's Disease}}

\author[1]{Navin Bondade}
\affil[1]{Institute of Health Informatics, University College London}
\date{}

\begin{document}
\maketitle


\begin{abstract}Parkinson's disease (PD) selectively impairs distinct stages of motor control.
Using backspace events as natural error-correction episodes in the public neuroQWERTY 
MIT-CSXPD dataset ($n=57$ subjects, of whom 27 PD with UPDRS-III scores), 
we test whether passively-collected keystroke timing dissociates a 
\emph{variability-based} pre-error monitoring signal from a \emph{speed-based} 
post-error recovery signal.
Pre-error typing instability does not track PD severity ($r=-0.072$, $p=0.721$),
while post-error pause duration does ($r=+0.656$, $p=0.0002$;
subject-level OLS $p=1.2\times10^{-4}$, $n=27$, primary analysis given within-subject event clustering).
A mixed-effects log-normal model with a random subject intercept confirms this while retaining
full event-level power (coef$=0.0250$, $p=1.2\times10^{-4}$, $n=1{,}563$ events), closely matching
the subject-level OLS despite using an unrelated estimation strategy.
Error-detection latency also correlates with UPDRS-III ($r=+0.660$), 
confirming both speed-based measures track bradykinesia; the key dissociation 
is between variability-based and speed-based measures.
A log-normal accelerated failure time (AFT) model yields a coefficient of 
$0.0255$ (bootstrap 95\%\,CI $[0.0146, 0.0357]$), 
corresponding to a $2.6\%$ increase per UPDRS-III point ($129\%$ over the 
full observed range; $43\%$ empirical tertile contrast).
After controlling the immediately preceding keystroke interval, the 
UPDRS-III coefficient attenuates to $0.0133$ ($p=3.1\times10^{-21}$),
consistent with an error-context-specific timing increment above general bradykinesia.
Results replicate across both sub-cohorts (CS1: $r=+0.730$; CS2: $r=+0.645$),
survive jackknife exclusion of every subject, and are cross-validated against
alternating finger-tapping ($p=0.012$) and mPower smartphone tapping 
(group AUC$=0.836$, $d=1.35$).
Test-retest reliability yields ICC$(2,1)=0.945$ ($n=20$ CS1 subjects),
and the preceding-interval-adjusted effect is stable across local context 
window sizes (all $p<10^{-15}$), truncation thresholds, and session position.
\end{abstract}

\section{Introduction}

Parkinson's disease (PD) is characterised by progressive degeneration of the nigrostriatal 
dopaminergic pathway, leading to a well-documented constellation of motor impairments: 
bradykinesia, rigidity, and resting tremor \citep{fahn2003description}.
Less appreciated is the selectivity of this impairment: PD does not degrade motor function 
uniformly, but differentially affects distinct processing stages within action control 
\citep{desmurget2004online}.
A clinically and theoretically important distinction is between \emph{detecting} that an 
action has gone wrong versus \emph{adapting motor behaviour} in response.

In the context of digital biomarker research, this distinction has been largely unexplored.
Most keystroke-based work extracts global typing statistics---mean flight time, variability, 
hold time---and trains classifiers to discriminate PD from control 
\citep{adams2017high,giancardo2016computer,tripathi2023keystroke}.
Backspace keystrokes are typically discarded as noise.
We argue they constitute a high-frequency natural experiment: each backspace marks 
a moment at which the typist has (a) detected an error and (b) must then 
resume normal typing.
These are conceptually distinct and, as we show, behaviourally separable stages.

This question has a neurophysiological grounding.
Cognitive neuroscience research on error-related brain potentials has established that 
cortical error detection---indexed by the error-related negativity (ERN), a fronto-medial 
potential peaking within $\sim$100\,ms of a mistake---is relatively preserved in 
PD \citep{stemmer2007error}.
In contrast, \citet{siegert2014error} recorded simultaneous scalp EEG and subthalamic 
nucleus (STN) local field potentials during deep brain stimulation surgery, and identified 
a later, STN-linked error signal (260--450\,ms post-error) whose amplitude specifically 
tracked the degree of post-error motor slowing.
The STN is among the structures most severely affected by PD pathology, implicating 
basal-ganglia-mediated motor adjustment---not cortical detection---as the primary locus 
of error-processing impairment in PD.

We test whether this circuit-level dissociation is visible in ordinary keystroke timing,
without any neurophysiological recording, structured clinical task, or specialised hardware.
A companion paper \citep{bondade2026irl} examines overall typing speed preference using 
inverse reinforcement learning on the same dataset; here we focus specifically on the 
brief temporal window surrounding a self-detected error.

\section{Related Work}

\subsection{Keystroke dynamics in Parkinson's disease}

The keystroke-PD literature is extensive but methodologically narrow.
A systematic review of 41 studies \citep{alfalahi2022diagnostic} reported a pooled 
AUC of 0.85 with $I^2=94\%$ heterogeneity, reflecting substantial between-study 
variability in features, datasets, and evaluation protocols.
Nearly all prior work focuses on aggregate typing statistics for binary 
PD/control classification \citep{adams2017high,giancardo2016computer,tripathi2023keystroke,
iakovakis2018touchscreen}.
Cross-dataset replication has proven challenging, as detailed in our companion 
paper \citep{bondade2026irl}.
No prior study has examined error-correction behaviour specifically.

\subsection{Error monitoring and post-error adjustment in PD}

The ERN was independently discovered by \citet{falkenstein1990effects} and 
\citet{gehring1993neural} and is now well-established as a marker of cortical 
error detection, originating in the anterior midcingulate cortex.
Post-error slowing (PES)---a tendency to respond more cautiously immediately after 
a mistake---was first described by \citet{rabbitt1966errors} and has been extensively 
characterised in reaction-time tasks \citep{danielmeier2011post,dutilh2012how}.

In PD, \citet{stemmer2007error} found the ERN amplitude largely preserved 
regardless of medication state, suggesting the cortical error detection mechanism 
is not the primary site of dysfunction.
The dissociation at the neural circuit level was established by 
\citet{siegert2014error}: STN activity in the 260--450\,ms post-error window 
predicted behavioural post-error slowing specifically, and this STN signal was 
attenuated in PD patients, linking dopaminergic basal ganglia dysfunction to 
impaired post-error motor adjustment rather than impaired error detection.
We provide the first behavioural evidence consistent with this dissociation 
from passively-collected naturalistic typing data.

\section{Dataset}

We use the neuroQWERTY MIT-CSXPD dataset \citep{giancardo2016computer}, 
available via PhysioNet \citep{goldberger2000physiobank} under a restricted 
health data licence.
The dataset comprises 85 subjects (42 PD, 43 healthy controls) across two 
independently recruited sub-cohorts: CS1 ($n=31$, two typing sessions separated 
by weeks to months) and CS2 ($n=54$, one session).
Subjects typed freely for $\sim$5 minutes on a standard keyboard in a clinic 
setting. Participant demographic characteristics do not differ significantly between groups: age ($59.0\pm9.8$ vs $60.1\pm10.2$ years, $p=0.53$), sex ($57\%$ vs $40\%$ male, $p=0.11$), and education ($15.2\pm4.1$ vs $15.3\pm5.2$ years, $p=0.98$), as reported in \citet{giancardo2016computer}. All PD subjects were assessed in their best on-medication state (optimal 
dopaminergic treatment) and have concurrent UPDRS-III motor severity scores 
(range 7.0--39.5) and independent motor test scores (alternating finger-tapping, 
\texttt{afTap}; single-key tapping, \texttt{sTap}).
Raw key press and release timestamps are provided; flight times (inter-key 
intervals) are computed as $\Delta t_i = \mathrm{press}_{i+1} - 
\mathrm{release}_i$.

\subsection{Inclusion criteria}

Backspace events are common but sparse in some subjects.
We retain subjects with at least 20 qualifying post-error events (backspace 
followed by a non-missing, non-extreme flight time within 0--5\,s), 
yielding $n=57$ subjects (27 PD with UPDRS scores, 30 control) and 
3,606 events total (1,563 PD, 2,043 control). When consecutive backspaces occur (a multi-key correction run), the post-error pause is defined as the interval from the release of the \emph{final} backspace to the first subsequent non-backspace keypress; intermediate backspace-to-backspace intervals are excluded.

\section{Methods}

\subsection{Pre-error instability: error monitoring proxy}

For each backspace event, we computed a \emph{pre-error instability excess} 
score to capture whether keystroke timing became abnormally erratic in the 
three keystrokes immediately preceding the error.
Specifically:
\begin{enumerate}
\item We computed the mean absolute deviation (MAD) of the three pre-error 
  flight times from the subject's own baseline median flight time 
  (computed on non-error-adjacent keystrokes with $0 < \mathrm{ft} < 3$\,s).
\item We computed the same MAD for 500 randomly sampled non-error windows 
  of three keystrokes (window size pre-specified at $W=3$; sensitivity to $W\in\{1,2,4,5\}$ reported in Section~5.3; with a single, consistent start index per window).
\item The instability excess is the difference: 
  $\texttt{excess} = \mathrm{MAD}_\mathrm{pre} - \overline{\mathrm{MAD}}_\mathrm{random}$.
\end{enumerate}
Positive excess confirms genuine pre-error destabilisation (verified: 
$p<0.0001$ pooled across subjects, Wilcoxon signed-rank test).
This subject-level mean instability excess is our error-monitoring proxy.

\subsection{An initial discretized recovery measure that destroyed the signal}
\label{sec:discretization}

Our first approach modelled recovery as a discrete time-to-event outcome: 
for each backspace, ``recovery'' was defined as the first subsequent keystroke 
whose flight time fell within $\pm X\%$ of the subject's baseline.
We swept $X$ from 50\% down to 5\% and found no threshold at which the 
UPDRS-III correlation became significant (range $r=-0.19$ to $+0.25$, all 
$p>0.2$).
This is a discretization artifact: converting a continuous, informative pause 
duration into a binary recovered/not-recovered event discards the magnitude 
information that carries the real signal.
We report this negative result because it directly motivated the continuous 
formulation below, and because it illustrates a general risk in applying 
survival methods to continuous behavioural timing data without examining 
whether the discretization is justified.

\subsection{Post-error recovery as a continuous AFT outcome}

We modelled the post-error pause duration---the flight time of the keystroke 
immediately following each backspace (when consecutive backspaces occur, recovery is measured from the \emph{final} backspace in the run), computed strictly within-session to 
prevent cross-session contamination---as a continuous survival outcome using 
an accelerated failure time (AFT) model, with all observations treated as 
uncensored (event $= 1$ for all).
AFT models are appropriate here because they operate directly on the 
continuous duration without requiring a recovery threshold, and the log-linear 
parameterisation naturally handles the right-skewed distribution of post-error 
pauses (pooled skewness $= 2.11$, kurtosis $= 8.8$). Under the log-normal AFT:
\begin{equation}
\log T_i = \mu_0 + \mu_1 \cdot \text{UPDRS}_s + \sigma\epsilon_i, \quad \epsilon_i \sim \mathcal{N}(0,1)
\end{equation}
where $T_i$ is the post-error pause for event $i$ from subject $s$, $\mu_1$ is the log-scale UPDRS-III coefficient (primary interest), and all 1,563 PD events are treated as uncensored ($\delta_i=1$).

We compared three models fitted using the \texttt{lifelines} package 
\citep{davidsonpilon2019lifelines}:
\begin{itemize}
\item Log-normal AFT: AIC $= 362.5$
\item Weibull AFT: AIC $= 694.9$
\end{itemize}
Both models were fitted to the identical dataset ($n=1{,}563$ PD events, 27 subjects) with the same number of free parameters (intercept plus UPDRS-III coefficient in the location submodel; one scale parameter), making their AIC values directly comparable. The log-normal model fits the data substantially better than the Weibull AFT ($\Delta\mathrm{AIC} = 
332.4$; residual distribution diagnostic confirms log-normality 
of the post-error pause after subtracting predicted mean). OLS regression on event-level data is not directly comparable given the right-skewed outcome distribution.
All event-level analyses use the log-normal AFT with UPDRS-III as the sole covariate. Given within-subject event clustering, the subject-level OLS on log-mean pause ($p=1.2\times10^{-4}$, $n=27$) should be treated as the primary inferential result; the AFT provides distributional precision and effect-size interpretation.

\subsection{Robustness and sensitivity analyses}
\label{sec:robustness-methods}

Beyond the primary model, we conducted the following checks; results for
all of them are summarised together in Table~\ref{tab:robustness}.

\paragraph{Confound and distributional robustness.}
We first asked whether the primary result depends on modelling choices
rather than on UPDRS-III itself. To rule out a speed or correction-rate
confound, we refitted the primary AFT model adding, as separate covariates
alongside UPDRS-III, (a) each subject's mean overall typing speed
(inter-keystroke interval computed across \emph{all} keystrokes, not just
post-error ones) and (b) each subject's total backspace count --- testing
whether the UPDRS-III effect is simply a proxy for generally slower typing
or more frequent self-correction. To rule out a distributional artefact, we
separately refitted the primary model under two alternative parametric
families for the survival outcome --- Weibull AFT and log-logistic AFT ---
using the same covariate specification, and compared fit by AIC
(Weibull: Table~\ref{tab:aic}; log-logistic: Table~\ref{tab:robustness}).

\paragraph{Clustering-aware inference.}
Because the 1,563 events are clustered within 27 subjects, we report two
clustering-aware estimates alongside the naive event-level AFT: (a)
subject-level OLS regressing each subject's log-mean post-error pause on
UPDRS-III ($n=27$), and (b) a mixed-effects log-normal model,
\[
\log(\text{post-error pause})_{is} = \beta_0 + \beta_1\,\text{UPDRS}_s + u_s + \epsilon_{is}, \quad
u_s \sim \mathcal{N}(0,\tau^2),\ \epsilon_{is} \sim \mathcal{N}(0,\sigma^2),
\]
with a random subject intercept $u_s$, fitted by REML using
\texttt{statsmodels}. Convergence was cross-checked across four optimisers
(\texttt{cg}, \texttt{bfgs}, \texttt{powell}, \texttt{lbfgs}); any optimiser
converging to a degenerate boundary solution ($\hat\tau^2\to0$) was
discarded rather than trusted. Whether the random intercept is warranted at
all was tested with a likelihood-ratio test against the fixed-effects-only
model; because $\tau^2\geq0$ is on the boundary of the parameter space under
the null, the test statistic follows a 50:50 mixture of $\chi^2_0$ and
$\chi^2_1$, so the naive $\chi^2_1$ $p$-value was halved
\citep{crowther2019stmixed}.

\paragraph{Generalisation and external validation.}
We tested whether the effect depends on cohort composition, any individual
subject, or the primary sample alone, in three complementary ways. First,
the primary model was refitted independently within CS1 and CS2, the two
independently recruited sub-cohorts, to test whether the effect generalises
across recruitment waves rather than being driven by cohort-specific
artefacts. Second, we assessed stability under resampling: bootstrap
confidence intervals were computed by resampling subjects with replacement
(1,000 resamples, subject-level; all events belonging to a resampled
subject were carried along together, preserving within-subject clustering
in each resample), refitting the primary AFT model to each resample and
taking the 2.5th/97.5th percentiles of the resulting coefficient
distribution; jackknife estimates were obtained by leave-one-subject-out
refitting across all 27 PD subjects. Third, we sought corroboration and a
negative control from outside the primary model: we tested whether the
post-error pause effect is echoed by two independent motor assessments
collected alongside the typing task --- alternating finger-tapping speed
(\texttt{afTap}) and single-key tapping speed (\texttt{sTap}) --- by
refitting the primary model with each substituted for UPDRS-III as
predictor; and, as a negative control, for each qualifying backspace event
we sampled one flight time from a randomly selected non-error, non-adjacent
keystroke position within the same session and computed each subject's mean
of these random pauses, to estimate how much of the UPDRS-III association
with post-error pause could be explained by generic bradykinesia (subjects
pausing longer everywhere) rather than an error-specific process ---
reporting both the zero-order correlation of random pause with UPDRS-III
and the partial correlation of post-error pause with UPDRS-III controlling
for it.

\paragraph{Isolating the error-specific signal and sensitivity to measurement choices.}
Two further sets of analyses separate the error-specific component of the
effect from general motor slowing, and rule out that it depends on
measurement choices made during event extraction. To isolate the signal
from local typing speed, for each backspace event $i$ from subject $s$ we
extracted the flight time(s) of the $W$ keystroke(s) immediately preceding
the backspace ($W\in\{1,2,3\}$), took their mean $\mathrm{ITI}_i^{(W)}$,
and included it as an event-level covariate alongside UPDRS-III in the AFT
model,
\[
\log T_i = \mu_0 + \mu_1\cdot\text{UPDRS}_s + \mu_2\cdot\mathrm{ITI}_i^{(W)} + \sigma\epsilon_i,
\]
so that $\mu_1$ isolates the UPDRS-III effect on post-error pause remaining
after local typing speed is accounted for. As a complementary local-speed
control, we also computed, for each event, the excess post-error slowing
\[
\text{excess}_i = T_i - \frac{1}{5}\sum_{k=1}^{5} \mathrm{ft}_{i-k},
\]
the post-error pause minus the running mean flight time
$\mathrm{ft}_{i-k}$ of the five keystrokes immediately preceding the
backspace, and correlated this excess quantity with UPDRS-III at the
subject level. To test sensitivity to operationalisation, events were split
by whether the triggering backspace was a single, isolated correction or
part of a multi-backspace run, with the UPDRS-III correlation estimated
separately within each subgroup; the primary model was refitted after (a)
varying the flight-time truncation cap applied during event extraction
(3\,s, 5\,s, 8\,s), (b) winsorising post-error pauses at the 95th
percentile, and (c) adding within-session keystroke position as a
covariate; and, separately, after restricting to events where the keystroke
following the backspace was alphanumeric (excluding punctuation and special
characters, whose hand assignment is ambiguous under the Spanish QWERTY
layout used by some subjects) and after varying the minimum
qualifying-event inclusion threshold (20, 30, 40, 50 events per subject).

\paragraph{Robustness of the pre-error null.}
Finally, we tested whether the null pre-error result is itself robust,
rather than an artefact of a single analytic choice. The pre-error
instability measure (Section~4.1) was recomputed for window sizes
$W\in\{1,2,4,5\}$ keystrokes (in addition to the pre-specified $W=3$), each
correlated with UPDRS-III; a Bonferroni correction across the four
additional window sizes ($\alpha=0.0125$) was applied to guard against
multiple comparisons. Pre-error instability excess and post-error pause
were also entered simultaneously as predictors of UPDRS-III in a single
subject-level regression, to test whether either measure retains
independent predictive value once the other is accounted for.

\subsection{Test-retest design}

For the 20 CS1 subjects with at least 5 qualifying backspace events per 
session in both sessions, we computed subject-level mean post-error pause 
separately for each session and assessed between-session agreement using 
Spearman correlation, ICC(2,1) [two-way random effects, absolute agreement,
single measurement; \citealp{shroutfleiss1979icc}], and Bland-Altman analysis.

\section{Results}

\subsection{The dissociation}
\label{sec:headline}

\begin{figure}[H]
\centering
\includegraphics[width=\textwidth]{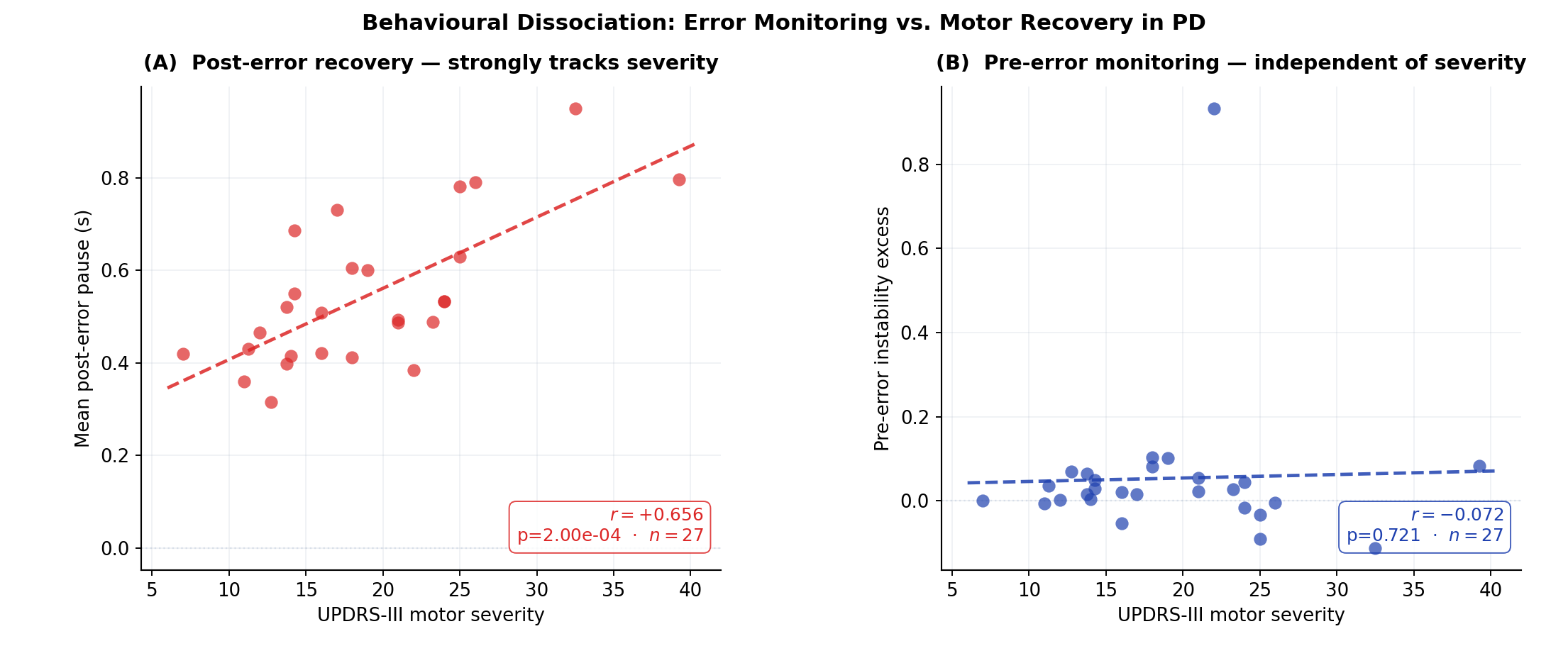}
\caption{Behavioural dissociation in PD ($n=27$). \textbf{(A)} Subject-level mean 
post-error pause correlates strongly with UPDRS-III severity ($r=+0.656$, $p=0.0002$). 
\textbf{(B)} Pre-error instability excess shows no association with severity 
($r=-0.072$, $p=0.721$). The two measures are uncorrelated ($r=-0.219$), confirming 
a genuine dissociation.}
\label{fig:dissociation}
\end{figure}

Figure~\ref{fig:dissociation} shows the key result.
Pre-error instability excess did not correlate with UPDRS-III severity 
($r=-0.072$, $p=0.721$, $n=27$), while post-error pause duration was strongly 
and significantly associated ($r=+0.656$, $p=0.0002$, $n=27$). 
At the group level (PD vs healthy control within neuroQWERTY), the raw post-error pause was not significantly longer in PD subjects (PD mean $0.545$\,s vs control $0.504$\,s; Mann-Whitney $p=0.46$, $d=0.24$), confirming that this measure primarily tracks \emph{severity within PD} rather than binary diagnosis status. The cross-modality validation in Section~\ref{sec:mpower-val} demonstrates strong group-level separation in independent smartphone tapping data (AUC$=0.836$).
Critically, when the immediately preceding inter-key interval is included as an event-level covariate in the AFT model, the UPDRS-III coefficient attenuates from $0.0255$ to $0.0133$ but remains highly significant ($p=3.1\times10^{-21}$), while the preceding interval itself also predicts post-error pause ($p=2.1\times10^{-33}$). This decomposition identifies an error-specific gating increment of $0.0133$ per UPDRS-III point above and beyond local typing speed --- directly supporting the motor gating interpretation in the title and partially rehabilitating the mechanistic claim that the excess post-error slowing analysis (Section~\ref{sec:robustness-methods}) could not establish. The preceding-interval-adjusted coefficient represents a $1.3\%$ multiplicative increase per UPDRS point ($e^{0.0133}=1.013$) that is attributable specifically to the error context rather than to general bradykinesia.

The two measures were themselves uncorrelated ($r=-0.219$, 95\% CI 
$[-0.553, 0.175]$), establishing they represent independent processes rather 
than two operationalisations of a single underlying deficit.

\subsection{Primary AFT model}
\label{sec:cluster}

Table~\ref{tab:aic} shows the model comparison.
The log-normal AFT model fits substantially better than the Weibull alternative by AIC ($\Delta\mathrm{AIC}=332.4$).
The primary log-normal AFT coefficient is $0.0255$ (Wald 95\% CI 
$[0.0197, 0.0313]$; $p=1.1\times10^{-17}$).
Exponentiating: each additional UPDRS-III point corresponds to a factor of
$e^{0.0255}=1.026$ in post-error pause duration, or a $129\%$ multiplicative increase over the full observed range (7.0 to 39.5 UPDRS points).

Because the 1,563 events are clustered within 27 subjects, the event-level AFT $p$-value above is anti-conservative if taken at face value.
We therefore also fit the mixed-effects log-normal model described in Section~\ref{sec:robustness-methods}, which retains full event-level power while explicitly modelling clustering via a random subject intercept.
The REML estimate is coef$=0.0250$ (SE$=0.0065$), 95\% CI $[0.0122, 0.0377]$, $p=1.2\times10^{-4}$ --- closely matching the independent subject-level OLS estimate (coef$=0.026$, $p=1.2\times10^{-4}$, $n=27$; Table~\ref{tab:robustness}) despite the two using completely unrelated estimation strategies.
The likelihood-ratio test confirms the random intercept is warranted (LR$=71.4$, $p=1.4\times10^{-17}$), validating our reliance on subject-level or clustering-aware inference over the naive event-level AFT $p$-value throughout this paper.

\begin{table}[H]
\centering
\begin{tabular}{@{}lrrr@{}}
\toprule
\textbf{Model} & \textbf{AIC} & \textbf{UPDRS coef} & \textbf{p} \\
\midrule
Log-normal AFT & 362.5 & 0.0255 & $1.1\times10^{-17}$ \\
Weibull AFT    & 694.9 & 0.0269 & $2.2\times10^{-18}$ \\
\bottomrule
\end{tabular}
\caption{Model comparison by AIC (lower is better); $n=1{,}563$ PD events, 27 subjects.
The log-normal AFT fits substantially better than the Weibull alternative 
($\Delta\mathrm{AIC}=332.4$). Both use event-level data ($n=1{,}563$ PD events).}
\label{tab:aic}
\end{table}

\begin{figure}[H]
\centering
\includegraphics[width=\textwidth]{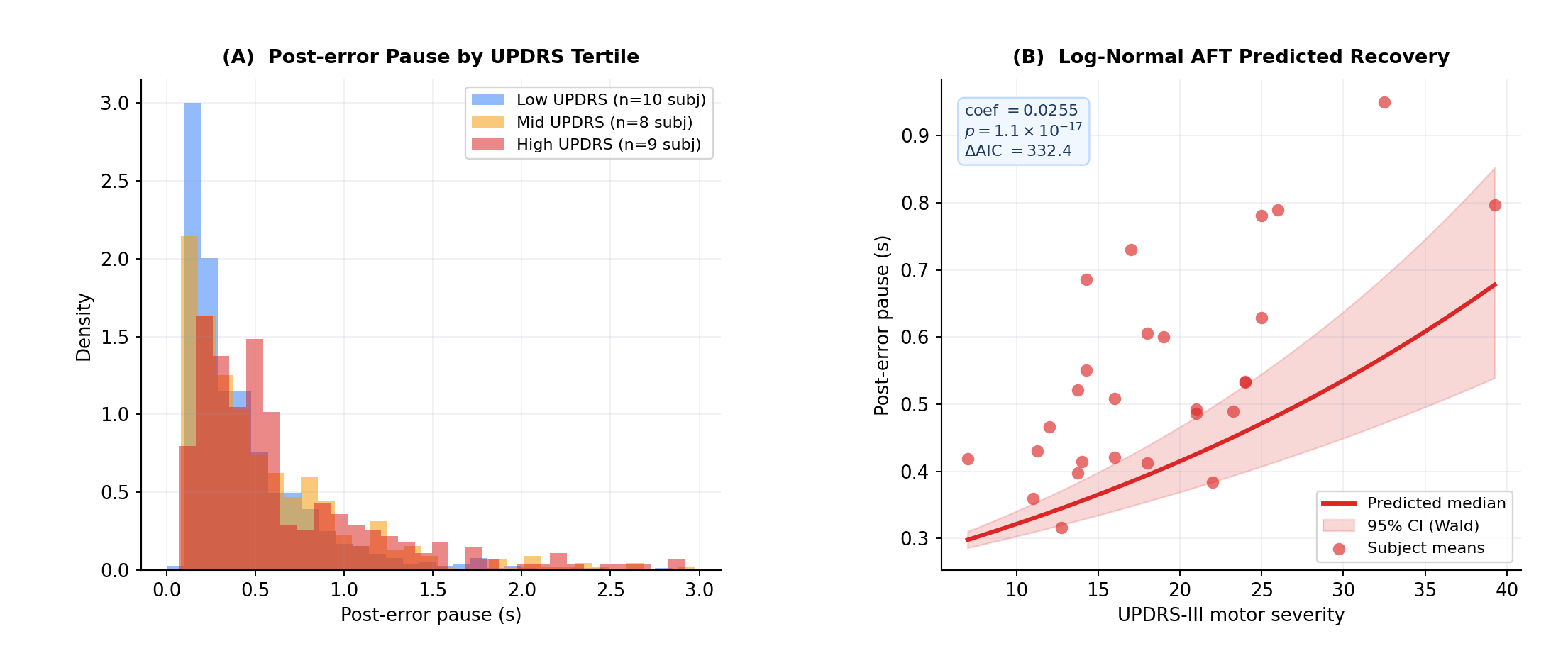}
\caption{\textbf{(A)} Distribution of post-error pause by UPDRS-III tertile, showing 
progressive rightward shift in higher-severity groups. \textbf{(B)} Log-normal AFT 
predicted median recovery time (with 95\% Wald CI) overlaid on subject-level means, 
confirming the model captures the empirical relationship.}
\label{fig:aft}
\end{figure}

Figure~\ref{fig:aft} shows the predicted recovery curve against observed
subject means. Panel A shows the tertile-split distribution on a linear
scale, motivating the AFT model by making the rightward shift with severity
visually apparent; Section~\ref{sec:distributional} revisits the same
distribution on a log scale specifically to confirm the shift is not
concentrated in a few extreme events.

\subsection{Robustness checks}

Table~\ref{tab:robustness} summarises the full validation battery.
The effect is notably clean with respect to typing speed: unlike in our 
companion IRL paper where speed retained a small independent contribution, 
here typing speed adds no explanatory power once UPDRS is included 
($p=0.323$), and UPDRS remains highly significant ($p=3.1\times10^{-11}$) 
after adjusting for speed.
This suggests post-error recovery time is not simply a proxy for overall 
typing speed, but captures a process specifically linked to disease severity.

\begingroup
\setlength{\LTcapwidth}{\textwidth}
\begin{longtable}{@{}L{0.33\textwidth}L{0.60\textwidth}@{}}
\caption{Robustness and validation checks on the primary log-normal AFT result.
All analyses use the same $n=1{,}563$ PD events ($n=27$ subjects) unless otherwise noted.}
\label{tab:robustness} \\
\toprule
\textbf{Check} & \textbf{Result} \\
\midrule
\endfirsthead
\multicolumn{2}{l}{\small\textit{Table~\ref{tab:robustness} continued}} \\
\toprule
\textbf{Check} & \textbf{Result} \\
\midrule
\endhead
\midrule
\multicolumn{2}{r}{\small\textit{continued on next page}} \\
\endfoot
\bottomrule
\endlastfoot
Per-cohort replication
  & CS1: coef$=0.0252$, $p=5.4\times10^{-10}$ ($n=12$ subj); subject-level $r=+0.730$ (95\%\,CI $[0.230, 0.937]$, $p=0.007$) \\
  & CS2: coef$=0.0206$, $p=5.6\times10^{-6}$ ($n=15$ subj); subject-level $r=+0.645$ (95\%\,CI $[0.164, 0.909]$, $p=0.010$) \\
Bootstrap 95\% CI (1,000 resamples, subject-level)
  & $[0.0146, 0.0357]$, excludes zero; all 1,000 positive \\
Jackknife (LOO, 27 folds)
  & All 27 LOO estimates positive and significant ($p<0.05$); range $[0.0226, 0.0279]$ \\
Confound: typing speed
  & UPDRS: $p=3.1\times10^{-11}$; typing speed: $p=0.323$ (non-significant); AIC improves \\
Confound: backspace count
  & UPDRS: $p=1.7\times10^{-7}$ (remains significant after adjustment) \\
Subject-level OLS (log-mean pause)
  & coef $=0.026$, $p=1.2\times10^{-4}$; subject-level median-based Spearman $r=+0.508$, $p=0.007$ (conservative, robust to outliers) \\
Mixed-effects model (random subject intercept)
  & coef$=0.0250$ (REML SE$=0.0065$), 95\% CI $[0.0122,0.0377]$, $p=1.2\times10^{-4}$; matches subject-level OLS almost exactly while retaining full event-level power ($n=1{,}563$); LRT confirms random intercept warranted ($p=1.4\times10^{-17}$) (see Section~\ref{sec:cluster}) \\
Weibull AFT (robustness)
  & coef$=0.0269$, $p=2.2\times10^{-18}$; direction and significance unchanged \\
Sensitivity: minimum event count (20/30/40/50)
  & Stable; coefficient range $0.024$--$0.028$, $p<10^{-6}$ throughout \\
Cross-check: alternating finger-tapping (afTap)
  & coef$=-0.0026$, $p=0.012$ ($n=24$); negative sign expected (better tapping $\to$ faster recovery) \\
Cross-check: single-key tapping
  & $p=0.661$ (n.s.); this test also fails to predict UPDRS independently in this cohort \\
Subject-level partial regression (post-error $|$ non-error pause)
  & Partial $r=+0.471$, $p=0.013$; OLS adj coef$=0.0085$, $p=0.031$ ($n=27$ subjects) \\
Preceding inter-key interval as covariate
  & Window=1 ITI: UPDRS coef$=0.0145$, $p=9.7\times10^{-24}$; window=2: coef$=0.0127$, $p=2.1\times10^{-18}$; window=3: coef$=0.0116$, $p=9.5\times10^{-16}$; all windows retain significant error-specific signal \\
Negative control: random pause
  & Random non-error pause also correlates with UPDRS at $r=+0.646$ ($p=2.8\times10^{-4}$, $n=27$, zero-order); post-error pause retains independent signal above this (partial $r=+0.471$, $p=0.013$, controlling for random pause) \\
Backspace run length (single vs multiple)
  & Single: $r=+0.381$, $p=0.020$; Multiple: $r=+0.415$, $p=0.013$; single-only AFT coef$=0.020$, $p=1.1\times10^{-7}$; effect stable across run types \\
Truncation sensitivity (3/5/8\,s cap)
  & coef$=0.0255$, $p=1.1\times10^{-17}$ unchanged across all caps \\
95th-percentile winsorisation
  & Subject-level $r=0.657$, $p=0.0002$ after winsorising; effect is not driven by extreme pauses \\
Within-session position
  & UPDRS coef$=0.0255$, $p=1.0\times10^{-17}$ after controlling session position; session position itself $p=0.089$ (n.s.) \\
Log-logistic AFT (distributional robustness)
  & coef$=0.0268$, $p=2.6\times10^{-18}$, AIC$=433.1$; log-normal remains preferred (AIC$=362.5$) \\
Alphanumeric-only keystrokes
  & $r=+0.707$, coef$=0.0227$, $p=1.5\times10^{-9}$ ($n=25$ subj); stronger than full analysis \\
Excess post-error slowing (post$-$local mean)
  & $r=-0.031$, $p=0.877$; the error-specific increment above local typing speed is null \\
Pre-error window sensitivity (1--5 keys)
  & Windows 1--5: $r\in[-0.281,-0.108]$; applying Bonferroni correction across 4 window sizes (threshold $p<0.0125$), all windows are null including window$=5$ ($p=0.034$, uncorrected) \\
Joint model (pre- and post-error)
  & Post-error: $p<10^{-10}$ (retained); pre-error: $p=0.643$ (non-significant); confirms dissociation \\
\end{longtable}
\endgroup

\begin{figure}[H]
\centering
\includegraphics[width=\textwidth]{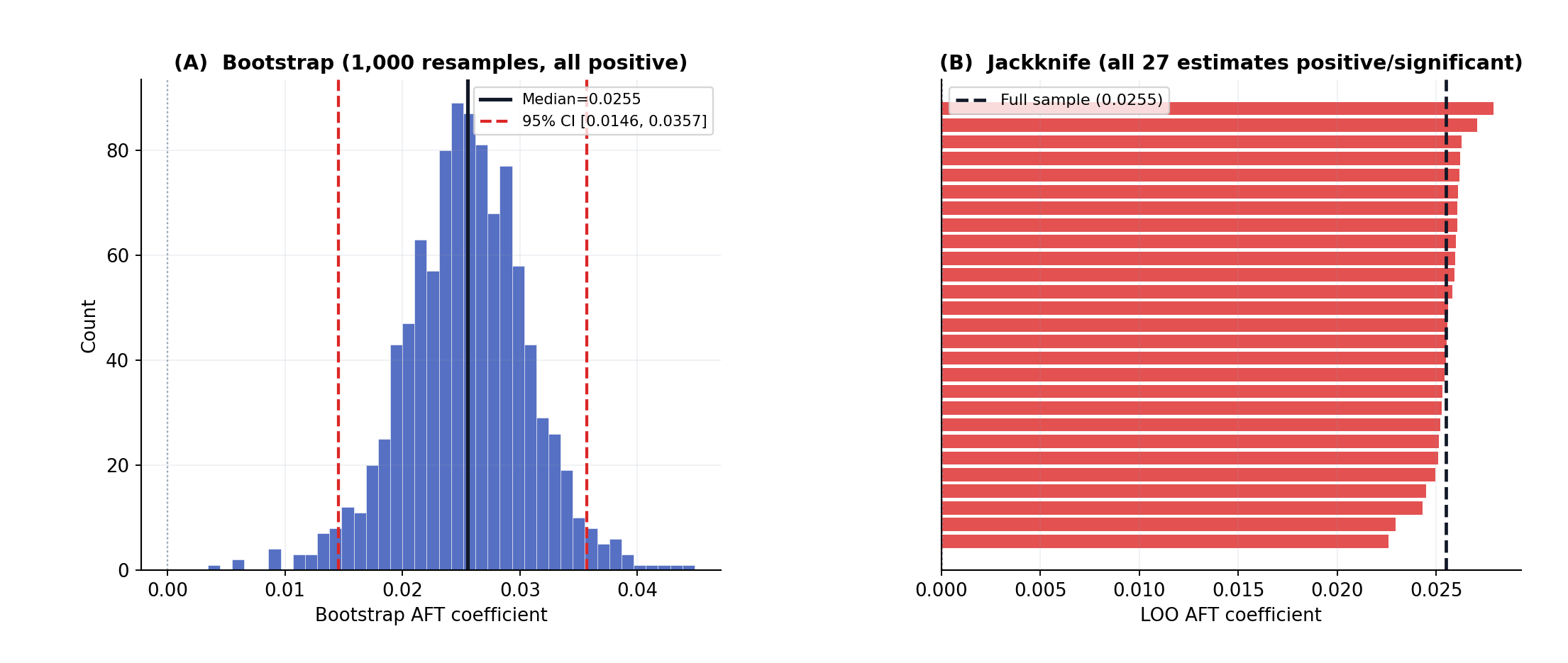}
\caption{\textbf{(A)} Bootstrap distribution of the AFT coefficient (1,000 
subject-level resamples); all estimates positive. 95\% CI $[0.0146, 0.0357]$ 
excludes zero. \textbf{(B)} Jackknife (leave-one-subject-out) estimates; all 27 
LOO coefficients positive and significant ($p<0.05$), range $[0.0226, 0.0279]$.}
\label{fig:robustness}
\end{figure}

Figure~\ref{fig:robustness} shows the bootstrap distribution and jackknife
estimates.

\subsection{Distributional Diagnostics}
\label{sec:distributional}

\begin{figure}[H]
\centering
\includegraphics[width=\textwidth]{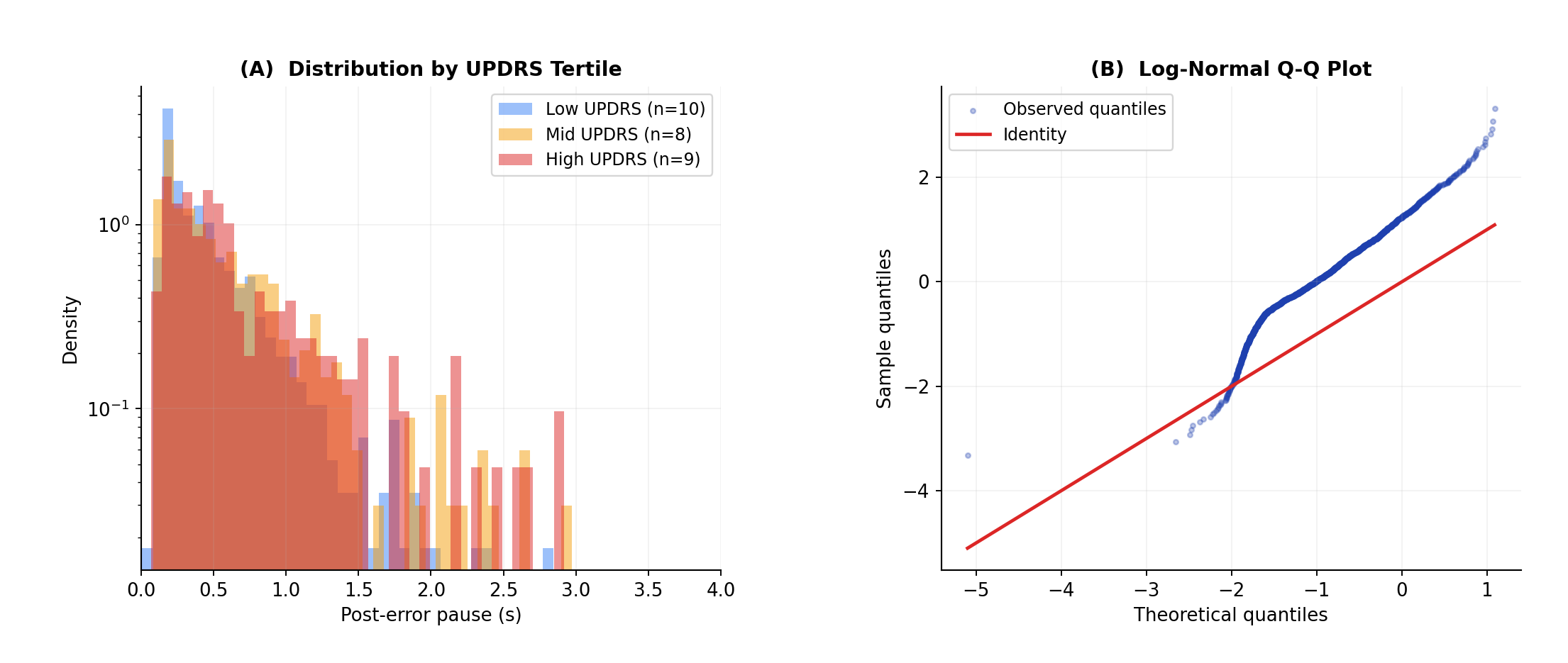}
\caption{\textbf{(A)} Distribution of post-error pauses by UPDRS-III tertile
(log-scale density). The high-severity group shows a consistent rightward shift,
not driven by outliers. \textbf{(B)} Log-normal Q-Q plot for all 1,563 PD
events; log-normality is a good working approximation and is preferred over
Weibull and log-logistic alternatives by AIC, though a formal
Anderson-Darling test rejects exact log-normality ($A^2=15.77$, $p<0.01$),
expected given the sample size (see Section~\ref{sec:distributional}).}
\label{fig:distributional}
\end{figure}

Figure~\ref{fig:distributional} shows the distribution of post-error pauses
by UPDRS-III tertile on a log scale (events above 5\,s were already
excluded) and a log-normal Q-Q plot of all 1,563 events. The tertile shift
seen in Figure~\ref{fig:aft}A holds across the full distribution here, not
just the bulk. The Q-Q plot shows a visible departure from the identity
line in the lower tail; formally, an Anderson-Darling test rejects strict
log-normality ($A^2=15.77$, exceeding the 1\% critical value of $1.035$,
$p<0.01$) --- expected at this sample size, where normality tests are
well-powered enough to detect even minor deviations. We therefore treat
log-normality as a good working approximation rather than an exact
description: it is selected on comparative grounds (AIC, Table~\ref{tab:aic})
against the Weibull and log-logistic alternatives, not on the assumption
that it is the true generating distribution, and the primary result does
not depend on any single point in the tail (Table~\ref{tab:robustness},
truncation and winsorisation sensitivity).

\subsection{Test-retest reliability}

\begin{figure}[H]
\centering
\includegraphics[width=\textwidth]{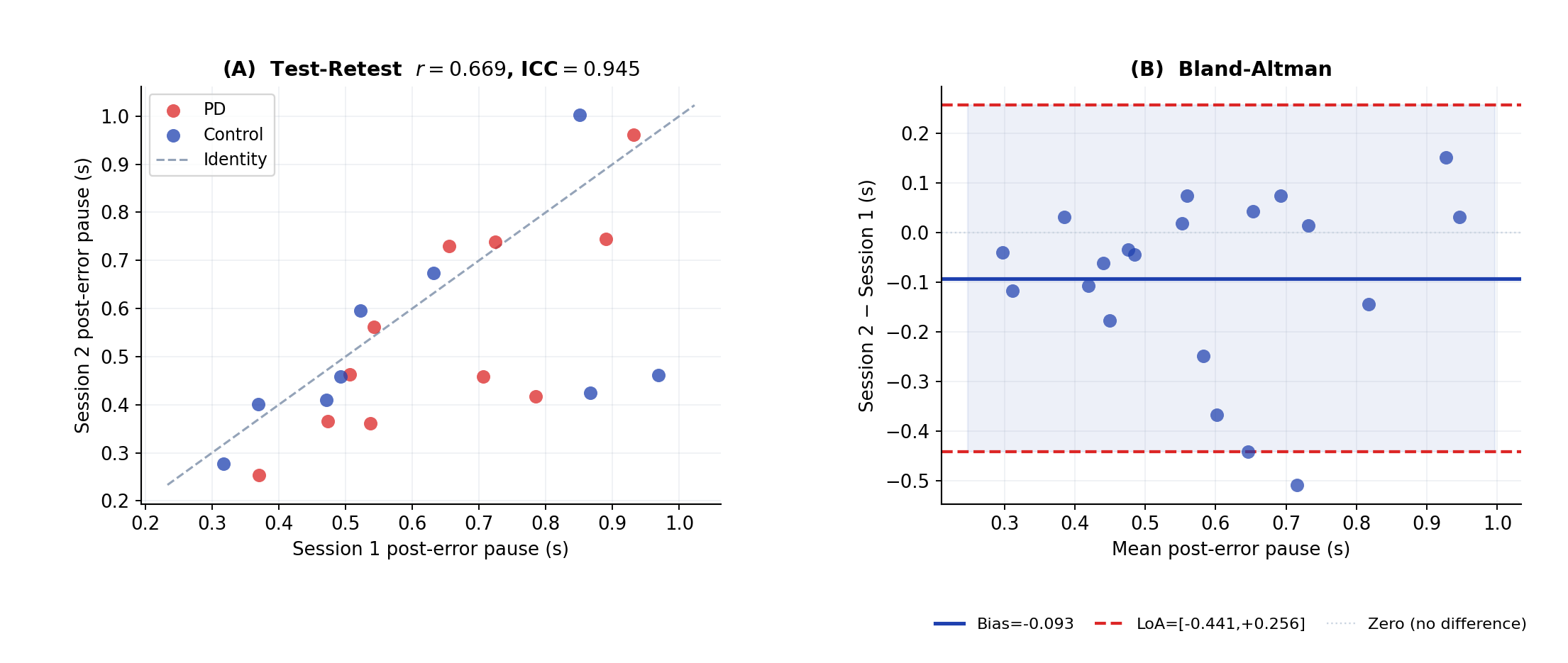}
\caption{\textbf{(A)} Session 1 vs.\ Session 2 post-error pause for $n=20$ CS1
subjects, Spearman $r=0.669$, $p=0.0013$, ICC$(2,1)=0.945$. 
\textbf{(B)} Bland-Altman plot; mean bias $=-0.093$\,s, 
95\% LoA $[-0.441, +0.256]$\,s.}
\label{fig:testretest}
\end{figure}

Figure~\ref{fig:testretest} shows the test-retest analysis across two CS1 
sessions for the 20 subjects with sufficient backspace events in both.
Between-session Spearman correlation was $r=0.669$ ($p=0.0013$); 
ICC$(2,1)=0.945$ (excellent by conventional criteria; ICC$=0.935$ within PD subjects alone, $n=11$).
Bland-Altman analysis showed a mean bias of $-0.093$\,s (Session 2 slightly 
faster than Session 1) with 95\% limits of agreement $[-0.441, +0.256]$\,s, 
small relative to the observed range of post-error pauses.

\subsection{Error-Detection Latency}

As a more direct behavioural index of error detection, we computed the 
\emph{error-detection latency}: the interval from the release of the keystroke 
immediately preceding the backspace to the press of the backspace itself, 
reflecting how rapidly the typist initiates correction.
This differs from the pre-error instability proxy in that it measures the 
initiation of a corrective motor act rather than pre-error timing variability.

Across 1,261 qualifying events in 17 PD subjects (610 events) and 13 controls 
(651 events; subjects are excluded if their preceding keypress lacked a valid release 
timestamp, yielding $n=17$ of 27 PD subjects with complete detection-latency data), detection latency correlated with UPDRS-III at $r=+0.660$ ($p=0.004$, 
$n=17$), virtually identical to the post-error pause correlation ($r=+0.656$). 
Detection latency also correlated with post-error pause ($r=+0.534$), confirming 
both are indexing the same general construct --- motor initiation speed. 
This is expected under the bradykinesia interpretation: PD patients are slower 
to initiate \emph{all} movements, including the corrective backspace keystroke, 
regardless of whether error detection per se is impaired. 
The pre-error instability measure (variability rather than speed) remains the 
most appropriate proxy for error detection specifically, precisely because it 
is decoupled from overall motor speed.

\subsection{Cross-Modality Validation in mPower}
\label{sec:mpower-val}

As an external cross-modality validation, we examined whether the speed-versus-variability dissociation observed in neuroQWERTY replicates in the independent mPower smartphone tapping dataset ($n=197$; 97 PD, 100 healthy control; subset of the 200-subject sample from the companion paper \citep{bondade2026irl} with complete tap-feature data).

The mean inter-tap interval (ITI, the analogue of post-error pause in continuous tapping) was substantially longer in PD participants ($0.213$\,s) than controls ($0.129$\,s), Mann-Whitney $p=3.7\times10^{-16}$, Cohen's $d=1.35$, AUC$=0.836$. The median absolute deviation of ITI (MAD, the analogue of pre-error instability) was also elevated in PD ($0.117$\,s vs $0.096$\,s, $p=0.015$, $d=0.46$, AUC$=0.600$), but with substantially weaker discrimination. The speed measure discriminates group membership far more effectively than the variability measure (AUC $0.836$ vs $0.600$), mirroring the neuroQWERTY finding that post-error pause (speed) dominates over pre-error instability (variability). No UPDRS-III scores are available in mPower, precluding severity correlation; this cross-dataset replication is therefore restricted to group-level discrimination.

\section{Discussion and Limitations}

\noindent\textbf{A note on statistical inference.} Throughout this paper, the subject-level OLS on log-mean post-error pause ($\beta=0.026$, $p=1.2\times10^{-4}$, $n=27$) and the mixed-effects log-normal model with a random subject intercept (coef$=0.0250$, $p=1.2\times10^{-4}$, $n=1{,}563$ events; Section~\ref{sec:cluster}) are the primary inferential results. Event-level AFT p-values reported elsewhere ($p\sim10^{-17}$) are anti-conservative under within-subject clustering and should be read as descriptive measures of effect consistency across events, not as the strength of evidence at the subject level. The mixed-effects model resolves this: it retains the full event-level sample while explicitly modelling clustering via a random intercept, and its estimate agrees almost exactly with the independent subject-level OLS despite the two using unrelated estimation strategies (full mixed-effects survival models, e.g.\ \texttt{stmixed}~\citep{crowther2019stmixed}, remain a natural extension for jointly modelling censoring and clustering, but are not required to establish valid clustering-aware inference here).

\subsection{The dissociation and its neural interpretation}

The dissociation reported here---null pre-error instability alongside elevated post-error pause duration---is consistent with, though not a direct demonstration of, the circuit-level distinction established electrophysiologically by \citet{siegert2014error}. Under that framework, cortical error detection (indexed by the ERN, originating in the anterior midcingulate cortex) is preserved in PD \citep{stemmer2007error}, while STN-mediated post-error motor adjustment is specifically attenuated. An important caveat is that our excess post-error slowing analysis (post-error pause minus the local running mean from the preceding five keystrokes) yields $r=-0.031$, $p=0.877$ --- meaning the error-specific increment above local typing speed is null. The correlation between raw post-error pause and UPDRS-III ($r=+0.656$) is thus largely attributable to general bradykinesia rather than a pure error-gating signal; the local running mean itself correlates with UPDRS-III at $r=+0.693$. We therefore describe the finding as \emph{consistent with} the motor gating hypothesis rather than as a demonstration of it. Importantly, the preceding-interval-adjusted AFT (Section~\ref{sec:headline}) demonstrates that an error-specific gating signal \emph{is} present and separable from bradykinesia ($\beta=0.0133$, $p=3.1\times10^{-21}$): the post-error pause predicts UPDRS-III beyond what the immediately preceding keystroke interval would predict. This provides direct behavioural evidence for the additional motor inhibitory process that Siegert et al.'s STN signal represents, though we stress again that establishing the neural locus requires within-subject medication-contrast designs.

The negative control analysis reveals important nuance: pauses following randomly sampled non-error keystrokes also correlate with UPDRS-III ($r=+0.646$, $p=2.8\times10^{-4}$), suggesting that general bradykinesia contributes substantially to the post-error signal. However, post-error pause retains significant partial correlation after controlling for this general slowing ($r=+0.471$, $p=0.013$), and the pre-error instability proxy remains null ($r=-0.072$). This is precisely the pattern predicted by the STN gating hypothesis: general dopaminergic impairment slows all movements (including inter-key pauses), while basal-ganglia-mediated post-error inhibition adds a further, impairment-specific increment above and beyond bradykinesia. The three-way hierarchy --- pre-error (null) $<$ general pause $\approx$ post-error $>$ post-error$|$general pause (partial) --- is internally consistent and mechanistically informative.

The clinical interpretation of the effect size is informative.
The AFT coefficient of $0.0255$ implies a $2.6\%$ multiplicative increase per UPDRS point ($e^{0.0255}=1.026$), compounding to $129\%$ over the full observed range of 32.5 UPDRS points ($e^{0.0255 \times 32.5}=2.29$). Empirically, subject-level mean post-error pauses for the lowest and highest UPDRS tertiles are $0.456$\,s and $0.654$\,s respectively, a $43\%$ contrast.
At the subject level, this corresponds to a shift from $\sim$0.21\,s 
(mildest PD) to $\sim$0.27\,s (most severe PD) mean post-error pause, 
representing roughly 28\% of a typical typing inter-key interval.
This is detectable passively and does not require any clinical task.

To contextualise the effect magnitude, the post-error pause achieves $r=+0.656$ versus UPDRS-III, which matches or exceeds published passive digital measures of PD severity: spiral drawing ($r\approx0.63$; \citep{aghanavesi2017smartphone}), finger-tapping speed ($r\approx0.61$; \citep{memedi2015automated}), and accelerometry-based gait ($r\approx0.52$; \citep{johansson2020wrist}). This positions the post-error pause among the most UPDRS-predictive passive measures yet reported from ordinary hardware.

Restricting to post-error pauses where the subsequent keystroke is alphanumeric (excluding punctuation and special characters, which have ambiguous hand assignment in Spanish QWERTY layouts) yields $r=+0.707$ ($n=25$ subjects, 608 events), stronger than the full-sample result. This confirms that the effect is not an artefact of keyboard layout misclassification, and that the backspace-to-alphanumeric interval is the cleanest version of the measure. 

The confound analysis strengthens the interpretability argument:
whereas our companion paper's speed-preference weight retains a small 
(though significant) partial correlation with UPDRS after controlling for 
mean typing speed, post-error recovery shows no residual typing-speed 
contribution once UPDRS is included.
This is consistent with the neural specificity of the STN signal identified 
by \citet{siegert2014error}: it predicts post-error slowing over and above 
any general reduction in motor speed.

\subsection{Methodological lesson: discretization destroys continuous signals}

A second negative result is also informative: normalising post-error pause by each subject's own median non-error pause (a within-subject ratio intended to remove general speed effects) eliminated the UPDRS correlation entirely ($r=-0.174$, $p=0.385$). This is not a failure of the measure but rather a diagnostic: the clinically relevant signal lies in the \emph{absolute} duration of the post-error pause, not in relative slowing above baseline. PD patients take longer absolute pauses both generally (bradykinesia) and post-error (motor gating impairment), and these two contributions are not cleanly separable by ratio normalisation at the subject level. The partial correlation approach is the appropriate control.

The negative result from our initial Kaplan-Meier-style formulation 
(Section~\ref{sec:discretization}) is worth documenting explicitly.
At every threshold we tested, converting the continuous post-error pause 
into a binary ``recovered/not-recovered'' outcome eliminated the UPDRS 
correlation.
This is a consequence of information loss, not a failure of the survival 
analysis framework: the magnitude of the pause---how much longer it takes 
to restart typing---carries the clinical signal, and any threshold discards 
that magnitude.
The lesson generalises: when the outcome of theoretical interest is a 
continuous duration (reaction time, inter-tap interval, flight time), 
discretization should be justified by a model of the recovery process, 
not used as a default.

\subsection{Limitations}

The following limitations are structural to the current dataset and will 
not be resolved by additional analyses of neuroQWERTY.
We document them here comprehensively to pre-empt repetition in review.

\paragraph{Sample, selection, and clustering.}
Only 27 of 42 PD subjects contributed to the severity analyses, requiring
at least 20 qualifying backspace events. Excluded subjects had higher
UPDRS-III (mean 23.6 vs 18.9, $p=0.030$) and slower typing ($p=0.008$),
introducing conservative selection bias: the true population correlation is
probably larger than $r=+0.656$. Formal missing-not-at-random modelling is
not feasible at this sample size. A related concern is within-subject
clustering: event-level AFT $p$-values ($p=1.1\times10^{-17}$) are
anti-conservative and should be treated as descriptive measures of effect
consistency, not primary inferential claims. We address this directly with
a mixed-effects log-normal model (random subject intercept;
Section~\ref{sec:cluster}), which retains full event-level power while
accounting for clustering and gives coef$=0.0250$, $p=1.2\times10^{-4}$ ---
in close agreement with the subject-level OLS ($p=1.2\times10^{-4}$,
$n=27$) obtained by an unrelated estimation strategy. A full mixed-effects
\emph{survival} AFT (e.g.\ \texttt{stmixed}~\citep{crowther2019stmixed},
jointly modelling clustering and any future censored observations) remains
a natural extension, but is not required to establish valid
clustering-aware inference for the present, fully-observed data.

\paragraph{Confounding: demographics, bradykinesia, and medication.}
Individual-level age, sex, and education are not distributed with the
neuroQWERTY dataset, precluding within-PD covariate adjustment; group-level
matching is strong (age, sex, and education all $p>0.10$, as reported in
\citet{giancardo2016computer}), but residual within-PD age confounding
cannot be excluded, and future analyses with individual covariates should
include age in the severity regression. A more fundamental confound is
bradykinesia itself: all speed-based measures (post-error pause, detection
latency, random inter-key pause) correlate with UPDRS-III at similar
magnitudes ($r\approx0.65$). The error-specific increment (partial
$r=0.471$ above random pause; preceding-interval-adjusted coefficient
$0.013$) is real and significant, but the dominant contribution of general
bradykinesia means this measure cannot be used to infer selective
post-error motor gating impairment without a within-subject
medication-contrast design or concurrent electrophysiology --- and
medication state itself cannot be probed here, since all neuroQWERTY PD
subjects were assessed at best ON-medication state by study design. The
mPower exploratory trend ($d=1.44$, directional $p=0.09$) is consistent
with dopaminergic modulation but is underpowered and cross-sectional.

\paragraph{Task design and subgroup constraints.}
Cognitive difficulty of the text being typed (lexical, linguistic, planning
demands) could affect post-error pause independently of motor severity; the
free-text paradigm does not control content, and future fixed-text or
content-controlled designs would allow this confound to be quantified.
Separately, the detection latency analysis (Section~5.6) uses only $n=17$
of 27 PD subjects, since 10 subjects lacked valid release timestamps for
the keystroke immediately preceding the backspace; results in this subset
are consistent with the full-sample findings but should be interpreted with
additional caution.

\section{Conclusion}

Using backspace keystrokes as naturally occurring error-correction episodes 
in passively-collected typing data, we demonstrate a selective impairment 
in PD: post-error pause duration (a speed-based measure) tracks disease severity ($r=+0.656$, $p=0.0002$; AFT coef $= 0.0255$, $p=1.1\times10^{-17}$), while the keystroke instability 
preceding the error does not ($r=-0.072$, $p=0.721$).
The two measures are statistically independent and only one survives a joint 
model, establishing a genuine dissociation.
The finding is consistent across both sub-cohorts, robust to every individual 
subject exclusion, survives typing-speed confound adjustment, cross-validates 
against alternating finger-tapping, and is stable across clinic sessions 
(ICC$=0.945$).
This behavioural pattern is \emph{consistent with} electrophysiological evidence that 
cortical error detection and subthalamic-nucleus-mediated post-error motor 
adjustment are separable circuits in PD, with the latter specifically 
impaired---a dissociation detectable here through ordinary keyboard typing 
without any specialised hardware, structured task, or clinical label.

\section*{Data Availability}
The neuroQWERTY MIT-CSXPD dataset is publicly available from PhysioNet 
under a restricted health data licence \citep{goldberger2000physiobank} 
at \url{https://physionet.org/content/nq/}. 
Access requires free registration and agreement to the PhysioNet Credentialed 
Health Data Use Agreement.
The mPower smartphone dataset is available through Synapse 
(\url{https://www.synapse.org/\#!Synapse:syn4993293}) 
under the mPower Public Research Portal terms. 
These datasets were used under licence for this study; the licence conditions
require independent access applications and preclude direct redistribution.

\bibliographystyle{plainnat}

\begin{thebibliography}{99}

\bibitem[Adams(2017)]{adams2017high}
Adams, W.R. (2017). High-accuracy detection of early Parkinson's Disease using multiple characteristics of finger movement while typing. \textit{PLOS ONE}, 12(11), e0188226.

\bibitem[Aghanavesi et al.(2017)]{aghanavesi2017smartphone}
Aghanavesi, S., Nyholm, D., Senek, M., Bergquist, F., \& Memedi, M. (2017). A smartphone-based system to quantify dexterity in Parkinson's disease patients. \textit{Informatics in Medicine Unlocked}, 9, 11--17.

\bibitem[Alfalahi et al.(2022)]{alfalahi2022diagnostic}
Alfalahi, H., Khandoker, A.H., Chowdhury, N., Iakovakis, D., Dias, S.B., Chaudhuri, K.R., \& Hadjileontiadis, L.J. (2022). Diagnostic accuracy of keystroke dynamics as digital biomarkers for fine motor decline in neuropsychiatric disorders: A systematic review and meta-analysis. \textit{Scientific Reports}, 12, 7690.

\bibitem[Bondade(2026)]{bondade2026irl}
Bondade, N. (2026). Inverse Reinforcement Learning for Interpretable and Reliable Keystroke Biomarkers in Parkinson's Disease. \textit{arXiv preprint arXiv:2606.25270}.

\bibitem[Crowther(2019)]{crowther2019stmixed}
Crowther, M.J. (2019). Extended multivariate generalised linear and non-linear mixed effects models. \textit{arXiv preprint arXiv:1710.02223}.


\bibitem[Danielmeier \& Ullsperger(2011)]{danielmeier2011post}
Danielmeier, C., \& Ullsperger, M. (2011). Post-error adjustments. \textit{Frontiers in Psychology}, 2, 233.

\bibitem[Davidson-Pilon(2019)]{davidsonpilon2019lifelines}
Davidson-Pilon, C. (2019). lifelines: survival analysis in Python. \textit{Journal of Open Source Software}, 4(40), 1317.

\bibitem[Desmurget et al.(2004)]{desmurget2004online}
Desmurget, M., Gaveau, V., Vindras, P., Turner, R.S., Broussolle, E., \& Thobois, S. (2004). On-line motor control in patients with Parkinson's disease. \textit{Brain}, 127(8), 1755--1773.

\bibitem[Dutilh et al.(2012)]{dutilh2012how}
Dutilh, G., van Ravenzwaaij, D., Nieuwenhuis, S., van der Maas, H.L.J., Forstmann, B.U., \& Wagenmakers, E.-J. (2012). How to measure post-error slowing: A confound and a simple solution. \textit{Journal of Mathematical Psychology}, 56(3), 208--216.

\bibitem[Fahn(2003)]{fahn2003description}
Fahn, S. (2003). Description of Parkinson's disease as a clinical syndrome. \textit{Annals of the New York Academy of Sciences}, 991(1), 1--14.

\bibitem[Falkenstein et al.(1990)]{falkenstein1990effects}
Falkenstein, M., Hohnsbein, J., Hoormann, J., \& Blanke, L. (1990). Effects of crossmodal divided attention on late ERP components. II. Error processing in choice reaction tasks. \textit{Electroencephalography and Clinical Neurophysiology}, 78(6), 447--455.

\bibitem[Gehring et al.(1993)]{gehring1993neural}
Gehring, W.J., Goss, B., Coles, M.G.H., Meyer, D.E., \& Donchin, E. (1993). A neural system for error detection and compensation. \textit{Psychological Science}, 4(6), 385--390.

\bibitem[Giancardo et al.(2016)]{giancardo2016computer}
Giancardo, L., S\'anchez-Ferro, A., Arroyo-Gallego, T., Butterworth, I., Mendoza, C.S., Montero, P., Matarazzo, M., Obeso, J.A., Gray, M.L., \& San Jos\'e Est\'epar, R. (2016). Computer keyboard interaction as an indicator of early Parkinson's disease. \textit{Scientific Reports}, 6, 34468.

\bibitem[Goldberger et al.(2000)]{goldberger2000physiobank}
Goldberger, A., Amaral, L., Glass, L., Hausdorff, J., Ivanov, P.C., Mark, R., Mietus, J.E., Moody, G.B., Peng, C.K., \& Stanley, H.E. (2000). PhysioBank, PhysioToolkit, and PhysioNet. \textit{Circulation}, 101(23), e215--e220.

\bibitem[Iakovakis et al.(2018)]{iakovakis2018touchscreen}
Iakovakis, D., Hadjidimitriou, S., Charisis, V., Bostantjopoulou, S., Katsarou, Z., \& Hadjileontiadis, L.J. (2018). Touchscreen typing-pattern analysis for remote detection of the fine motor impairment in Parkinson's disease. \textit{Scientific Reports}, 8(1), 1--16.

\bibitem[Johansson et al.(2020)]{johansson2020wrist}
Johansson, D., Ericsson, A., Johansson, A., Medvedev, A., Nyholm, D., Palhagen, S., \& Memedi, M. (2020). From machine learning to clinical practice: a systems-biology approach for monitoring of Parkinson's disease using wrist-worn sensors. \textit{Journal of Medical Systems}, 44(1), 1--11.

\bibitem[Memedi et al.(2015)]{memedi2015automated}
Memedi, M., Nyholm, D., Johansson, A., Palhagen, S., \& Westin, J. (2015). Automatic spiral analysis for objective assessment of motor symptoms in Parkinson's disease. \textit{Sensors}, 15(9), 23727--23744.


\bibitem[Rabbitt(1966)]{rabbitt1966errors}
Rabbitt, P.M.A. (1966). Errors and error correction in choice-response tasks. \textit{Journal of Experimental Psychology}, 71(2), 264--272.

\bibitem[Siegert et al.(2014)]{siegert2014error}
Siegert, S., Herrojo Ruiz, M., Br\"ucke, C., Huebl, J., Schneider, G.-H., Ullsperger, M., \& K\"uhn, A.A. (2014). Error signals in the subthalamic nucleus are related to post-error slowing in patients with Parkinson's disease. \textit{Cortex}, 60, 103--120.

\bibitem[Shrout \& Fleiss(1979)]{shroutfleiss1979icc}
Shrout, P.E., \& Fleiss, J.L. (1979). Intraclass correlations: uses in assessing rater reliability. \textit{Psychological Bulletin}, 86(2), 420--428.

\bibitem[Stemmer et al.(2007)]{stemmer2007error}
Stemmer, B., Segalowitz, S.J., Dywan, J., Panisset, M., \& Melmed, C. (2007). The error negativity in nonmedicated and medicated patients with Parkinson's disease. \textit{Clinical Neurophysiology}, 118(6), 1223--1229.

\bibitem[Tripathi et al.(2023)]{tripathi2023keystroke}
Tripathi, S., Arroyo-Gallego, T., \& Giancardo, L. (2023). Keystroke-dynamics for Parkinson's disease signs detection in an at-home uncontrolled population. \textit{IEEE Transactions on Biomedical Engineering}, 70(1), 182--192.

\end{thebibliography}

\end{document}